**Myriem ALIJO, PhD Candidate**
E-mail: alijomyriem90@gmail.com
**Professor Otman ABDOUN**
E-mail: abdoun.otman@gmail.com
**AbdelmalekEssaadi University, Polydisciplinary Faculty,Larache Morocco**
**Mostafa BACHRAN, PhD**
E-mail: bachrane@gmail.com
**Mohamed 5 University, LASTIMI EST Sale, Morocco**
**Professor Amal BERGAM**
E-mail: bergamamal11@gmail.com
**AbdelmalekEssaadi University, Polydisciplinary Faculty,Larache Morocco**

# OPTIMIZATION BY HYBRIDIZATION OF A GENETIC ALGORITHM WITH THE PROMOTHEE METHOD: MANAGEMENT OF MULTICRITERIA LOCALIZATION

***Abstract.*** *The decision to locate an economic activity of one or several countries is made taking into account numerous parameters and criteria. Several studies have been carried out in this field, but they generally use information in a reduced context. The majority are based solely on parameters, using traditional methods which often lead to unsatisfactory solutions.This work consists in hybridizing through genetic algorithms, economic intelligence (EI) and multicriteria analysis methods (MCA) to improve the decisions of territorial localization. The purpose is to lead the company to locate its activity in the place that would allow it a competitive advantage. This work also consists of identifying all the parameters that can influence the decision of the economic actors and equipping them with tools using all the national and international data available to lead to a mapping of countries, regions or departments favorable to the location. Throughout our research, we have as a goal the realization of a hybrid conceptual model of economic intelligence based on multicriteria on with genetic algorithms in order to optimize the decisions of localization, in this perspective we opted for the method of PROMETHEE (Preference Ranking Organization for Method of Enrichment Evaluation), which has made it possible to obtain the best compromise between the various visions and various points of view.*

Myriem Alijo, Otman Abdoun, Mostafa Bachran, Amal Bergam
_________________________________________________________________

### Introduction

Several scientific disciplines (economics, mathematical programming, statistics, etc.) are based on the idea of the existence of an optimal decision based on an objective criterion, seek to optimize, involves a criterion-based unique resolution approach, however this proves absurd, because several visions and sources of information have to be taken into account in order to assess the consequences of many actions, it is at this level that EI intervenes in order not to choose as a favorite a different action that gives rise to erroneous decisions.

The aim of designing an EI model for the location of future investments is based on a maximization approach to profit functions: investors maximize their profit function and choose to locate in the region where the expectation Profit is the highest.

The majority of methods used for optimal location are based on the analysis of the distribution of potential customers in a given area. Among the theoretical methods have found gravity models, the interactive model of competition that seek to locate points of sale as close as possible to customers in order to attract their greater part. While localization models try to find the closest sites to customers through mathematical algorithms [1] [2] [3].

In this paper, we seek to hybridize through genetic algorithms and multicriteria analysis methods (MA) to improve the decisions of territorial localization. The purpose is to lead the company to locate its activity in the place that would allow it a competitive advantage. This work also consists of identifying all the parameters that can influence the decision of the economic actors and provide them with tools using all the national and international data available to lead to a cartography and classification of favorable countries, regions or departments to the location.

### 1. Mathematical formulation of the problem

The problem to be solved is to simultaneously optimize the number and the territorial location of the investments and determine the capacities offered by each location area; In the case of our studies we have two objective functions:

The first concerns maximizing profits:

$Max\ E\ [Rp] = P n_{i=1} w_i E[Rp]$ (1)
$Subject\ to\ \sigma_{p2} = P_{n\ i=1}\ P_{n\ j=1}\ w_i\ w_j \sigma_{ij} \sigma_i \sigma_j = cte$ (2)
$P n_{i=1}\ w_i = 1$ (3)
$w_i \geq 0,\ i = 1, 2, ..., n$ (4)

And the second is to minimize risks:

$Min\ \sigma 2p = P n_{i=1} P n_{j=1} w_i\ w_j \sigma_{ij} \sigma_i \sigma_j$
$Subject\ to\ E\ [Rp] = P n_{i=1} w_i E[Rp] = R*$

*P$n_{i=1}$ wi= 1*
*wi ≥ 0 pour i = 1, 2... n*

With *wi* is the weighting of the company's action i.

### 1.1 Selection conditions:
We formulated questions in the form of a maintenance guide which we asked the three economic actors responsible for promoting investment in the Moroccan Regional Investment Center (CRI) in Casablanca, Kenitra and Tangier. The objective is to identify indicators that make localization decision-making more efficient in a country.

Companies that pursue to locate in a country collect information in the form of indicators, to evaluate possible scenarios and to clarify their choice of location; this modeling takes into account a series of criteria belonging to a defined category that we present hereinafter:

**Infrastructure Criteria:** Infrastructure is a prerequisite for investment attraction and success; it is supposed to be a direct factor in locating new businesses, but also an indirect factor in the formation of an economic space.

**Consumption of electricity:** The electricity consumption indicator measures the output of power stations over the total population.

- **Overall logistics performance index:** The overall score for the logistics performance index reflects a country's logistics perceptions based on the effectiveness of customs clearance processes, the quality of commercial infrastructure and related transport infrastructure.

**2.1.2. Economic Criteria:** Investors' perceptions of the investment climate help companies assess the degree of investment viability and the conditions of fair competition. There are many criteria to consider, we quote:

- **Inflation of GDP:** is the loss of the purchasing power of the currency which results in a general and lasting increase in prices, the loss of value of currency units is a phenomenon that strikes the national economy.
- **Domestic credit provided by the banking sector:** Investors, individuals or companies, often use banks to solicit financing and carry out their investments.
- **GDP growth:** The Gross Domestic Product (GDP) is the sum of value added within a country by all industries for a given

period, this indicator is often used when making short- and medium-term forecasts of a country's economic situation.

**2.1.3. Social Criteria:** The accumulation of human capital is a factor that largely determines the earnings capacity and employment prospects of individuals, and hence the level and distribution of income.

- **Unemployment:** refers to the labor force that is unemployed but is available or looking for work. It has a negative effect on investment when it exceeds 10% of the labor force
- **Population growth:** corresponds to the exponential growth rate of the population in the middle of year *n-1* to *n*,

**2.1.4. The administrative criteria:** This category makes it possible to make evaluations of the administrative services provided by the country.

- **Time required starting a business (days):** The time needed to set up a company is the number of calendar days required to complete all the procedures for legally operating a company.
- **Time required preparing and pay taxes (hours):** There are three main types of taxes that each investor is required to pay, business taxes, sales tax, or sales tax.

**2.1.5. Political criteria:** The attractiveness policies put in place by the Mediterranean countries reflect the willingness of the state to guarantee an attractive environment for investment, a source of local development and growth, and aim to protect and promote foreign investment through Rights and obligations.

- **Political stability:** Political stability in its entirety rests on two complementary and not-dissociable notions: the notion of national security and the notion of public security, which concerns citizens and their protection in all circumstances, the integrity of the territory of a state, protection of its population and the preservation of its national interests against all types of external threats and aggressions.

**1.2 Data development and model construction:** The decision to locate an economic activity in one or more countries is made taking into account many parameters and criteria. The new location trend is based on informational data, several studies have been carried out in this field, but they generally use information in a reduced context. The majority are based solely on socio-economic and / or political parameters, using traditional methods, which often lead to unsatisfactory solutions.

Optimization by Hybridization of a Genetic Algorithm with the PROMOTHEE
Method: Management of Multicriteria Localization
_______________________________________________________________

**Table 1: Presentation of the selection conditions and their notations**

| Category | Indicators | Notation | Conditions | Min/Ma |
|---|---|---|---|---|
| Infrastructure | Consumption of electricity | $C_{Infra}1$ | $C_{Infra}1$ R | Max |
| | Logistics Performance | $C_{Infra}2$ | $C_{Infra}2[1, 5]$ | Max |
| Economic | Inflation of GDP | $C_{Econ}1$ | $C_{Econ}1 \geq 4$ | Min |
| | Credit provided by bank | $C_{Econ}2$ | $C_{Econ}2$ R | Max |
| | GDP growth | $C_{Econ}3$ | $C_{Econ}3[3, 10]$ | Max |
| Social | Unemployment | $C_{Soc}1$ | $C_{Soc}1[1, 9]$ | Min |
| | Population growth | $C_{Soc}2$ | $C_{Soc}2[1, 4]$ | Max |
| Administrative | Start a business | $C_{Admi}1$ | $C_{Admi}1[6d, 10d]$ | Min |
| | Paying taxes | $C_{Admi}2$ | $C_{Admi}2[120h, 240h]$ | Min |
| Policy | Political stability | $C_{poli}$ | $C_{poli}[2, 5, -2, 5]$ | Max |

## 2. Methods of resolution

Two methods are commonly used: The ELECTRE methods (Elimination and Choice Translating Reality) initiated by Bernard Roy[5] and PROMETHEE (Preference Ranking Organization METHOD for Enrichment Evaluations) methods initiated by Jean Pierre Brans, Philippe Vincke and Bertrand Mareschal. 1986. [6]

**2.1 ELECTRE methods:** is a method of decision support system based on multiple criteria derived from Europe around 1960. Can be used in the assessment and classification according to strengths and weaknesses by a pairwise comparison to the same criteria. The purpose underlying the description of this method is rather theoretical and pedagogical. The method is very simple and it should be applied only when all the criteria have been coded in numerical scales with identical ranges. In such a situation we can assert that an action "a outranksb" (that is, "a is at least as good as b") denoted by aSb, only when two conditions hold.

On the one hand, the strength of the matching coalition must be strong enough to support the above statement. We mean the sum of the weights associated with the criteria forming this coalition. It can be defined by the following concordance index P (assuming, for the sake of simplicity of the formulas, that $\Sigma j \in J w j = 1$, where J is the set of the indices of the criteria):$c(aSb) = \Sigma_{\{j: gj(a) \geq gj(b)\}} wj$ *(where {j : gj (a) ≥ gj (b)}*

It is the set of indices for all criteria belonging to the coalition that match the aSb over classification relationship. the value of the concordance index must be greater than or equal to a given level of concordance, s, whose value is generally in the range *[0.5, 1 - min j∈J wj ], i. e., c(aSb) ≥ s*. On the other hand, no discordance against the assertion "a is at least as good as *b*" may occur. The discordance is measured by a discordance level defined as follows:

$d(aSb) = max_{j\,:\,g_j(a) < \,:\, g_j(b)} \{g_j(b) - g_j(a)\}.$

This level measures in a way the power of the discordant coalition, which means that if its value exceeds a given level, v, the statement is no longer valid. It exercises no power each time $d(aSb) \leq v$. The concordance and discordance indices must be calculated for each pair of shares (a, b) in the set A, where $a \neq b$. It is easy to see that such a calculation procedure leads to a binary relationship in complete terms in set A. Thus for each pair of actions *(a, b)*, only one of the following situations can occur:
- aSb and not bSa, i.e., aP b (a is strictly preferred to b).
- bSa and not aSb, i.e., bP a (b is strictly preferred to a).
- aSb and bSa, i.e., aIb (a is indifferent to b).
- Not aSb and not bSa, i.e., aRb (a is incomparable to b).

**2.2 PROMETHEE:**

The Preference Ranking Organization method METHod for Enrichment Evaluation (PROMETHEE) is a method developed by researcher BRANS JP in 1986. This method is used to determine relationships of over-ranking, incomparability and indifference between two scenarios better and worse, one assigns a score to each criterion, the objective is to evaluate the preference index of one scenario over the other. This index is then used to measure the attractiveness of a scenario against to another and the submission of this scenario in relation to all the others. [8] [9]

The PROMETHEE methods are based on an extension of the notion of criterion by the introduction of a function expressing the decision-maker's preference for one action over another, for each criterion, the decision-maker is called upon to choose one of the six forms curves shown below.
- The function I: is used when the data has a discrete character such as an ordinal classification or an all or nothing type value.
- Function II: is used when the indifference thresholds are clearly apparent in the data of the problem posed.
- Function III: is used when the data is such that the deviations between it is of a continuous character, or when all the intermediate values between the maximum and minimum values of these deviations are possible.
- Function IV: It is sometimes used when it can be asserted that a candidate is not at the same time neither strictly preferred to another, nor indifferent, this candidate Deviation from another will be assigned 1/2 point.
- The function V: is used when the indifference and strict preference thresholds are clearly apparent in the data of the multicriteria problem posed.

Optimization by Hybridization of a Genetic Algorithm with the PROMOTHEE
Method: Management of Multicriteria Localization
_______________________________________________________________

- Function VI (Gaussian distribution): is the most widely used function in Applications and is particularly appropriate in the case of a number of candidates sufficiently high, in this case the standard deviation of this distribution should be calculated.

If we take PROMETHEE methods, they represent themselves through following steps:

**Step 1:** For each criterion, one of the six forms of curves proposed in PROMETHEE as well as the parameters associated with it.

**Step 2:** For each action pair (*ai, ak*); the overall preference (degree of upgrade) as follows: $P(ai, ak) = \sum_{j=1}^{n} = \pi j . Fj(ai, ak)$

**Step 3:** Calculate incoming and outgoing flows for each action (*ai*),

$\varphi+ (ai) = P\ ak \in A, ai, ak P(ai, ak)$ Positive flux that expresses the force (*ai*) outflow;

$\varphi- (ai) = P\ ak \in A, ai, ak P(ak, ai)$ Negative flux that expresses the weakness of (*ai*) incoming flow

**Step 4:** Determine the 2 pre-orders total and proceed with the stocking of the actions:

The first total pre-order is to rank the shares in descending order of $\varphi+$

The second total preorder is to rank the actions in the ascending order of $\varphi-$
This method consists of storing the actions in descending order of the scores $\varphi(ai)$ defined as follow: $\varphi(ai) = \varphi+ (ai) - \varphi- (ai)$.

### 2.2.1 Multi-criteria preference index

The multicriteria preference index provides the degree of preference of the decision maker for one action over another while considering the set of all criteria defines a preference relation evaluated on I.

If (a, b) is close to 0 (respectively 1) we are in the presence of a weak (or strong) preference of an over b over all the criteria, this preference relation will be exploited in the different methods of the PROMETHEE family through upgrade flows. [10 - 12]

**3.2.2 Upgrade:** For each action pair *a* and *b* belonging to *I*, the values (a, b) and (b, a) are calculated; in this way a relation of the upgrade and built on *A*.

Let's look at how each action belongs to *I* is relative to the *n-1* other actions. This allows us to define two upgrade streams.

**3.2.3 Upgrade flow:** Multi-criteria flows are the linear combination of uni-criteria flows. The incoming and outgoing flows are introduced to allow the construction of a partial pre-order on all the actions by accepting that actions are incomparable (same performance).

Myriem Alijo, Otman Abdoun, Mostafa Bachran, Amal Bergam
_______________________________________________________________

**The outgoing flow** Φ + expresses how much one action has outclassed all others; the higher the flow, the better the action. The flow represents the outclass character of action a. $\Phi+ (a) = \frac{1}{n-1}\sum_{j=1}^{n} \pi(a, x_j)$

**The incoming flow** Φ- expresses how much an action *a* is outclassed by all other actions. The lower the flow and the better the action. The negative flow represents the outclassed character of the action. $\Phi- (a) = \frac{1}{n-1}\sum_{j=1}^{n} \pi(x_j, a)$

**The net flow** is the difference between the two outgoing and incoming flows; it makes it possible to carry out a total stocking of the shares. Information is less rich than partial storage, but it has the advantage of providing a complete ranking, which is often required in order to negotiate for decision-making.

Net flows are positive if the action is generally preferred to the other shares and negative if the other shares are on average preferred to the action. $\Phi(a) = \Phi+ (a) - \Phi- (a)$

### 3. Proposed Method

Technically, the complexity of an optimization problem is measured by its resolution problem, itself measured by the time required for an information processing system to execute a computer program which gives the solution. It is clear that the calculation time depends on the size of the data problem. [10].

The problem considered is a problem of choice of location of investments, this kind of combinatorial optimization problems is classified NP-complete. [13] It implies Non-polynomial increase of the space size of the solutions at the time of resolution, in particular when the number of nodes is multiple. It is understood that the use of a heuristic or metaheuristic algorithm does not guarantee the optimum solution, but generally leads to a good level of performance in terms of solution quality and response time. [14].

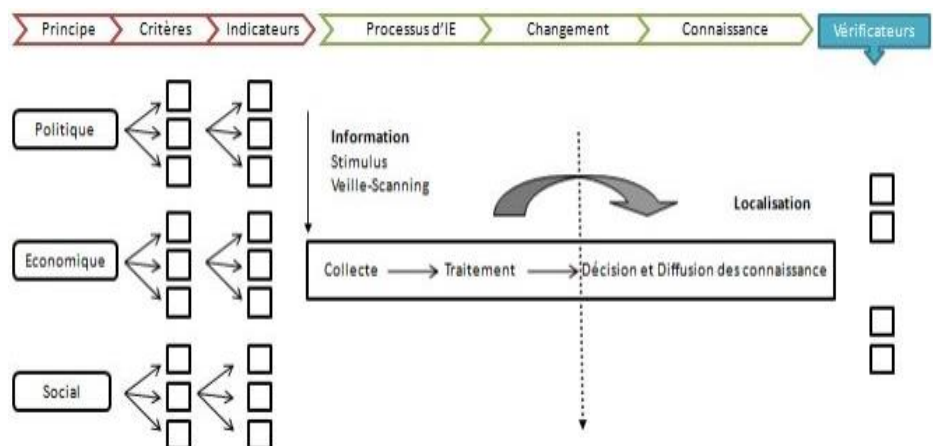

**Figure 1: Conceptual model of multi-criteria economic intelligence**

Optimization by Hybridization of a Genetic Algorithm with the PROMOTHEE
Method: Management of Multicriteria Localization
____________________________________________________________________

The proposed method is based on a hybrid genetic algorithm with a multicriteria decision support method, using a binary coding system, it incorporates a modification of the PROMETHEE method which classifies the actions in descending order, according to the performance target population, and then the initial population of solutions is calculated according to the centers derived from clusters. Uses an approach to the genetic algorithm to solve the problem. The algorithm uses binary coding with an adaptation of genetic operators. The caching technique is also implemented in the GA in order to improve its efficiency.

### 3.1 Resolution of problem:

By consulting the literature, we have been able to choose the genetic algorithm that is prepared to be best suited to such a case. Thus, an adaptation of the algorithm to solve the problem. We consider $I$, the whole of the countries of the Mediterranean, the reference point of our problem decision, crossing infrastructure criteria, economic criteria, social criteria, administrative criteria and criteria Policies; Defined in the set $J$ to From which the consequences are assessed $\sigma ij(i=1 and j=1)$ with i ∈ I and j ∈J.
I and J in extension are such that:
$I$ = *Algeria, Egypt, Spain, France, Italy, Libya, Morocco, Syria, Tunisia, Turkey*
$J$ = *Electricity consumption; Logistics performance index; Inflation, GDP deflator; Domestic credit provided by the banking sector, GDP growth, Total unemployment, Population growth, Time needed to start a business, Time to prepare and pay taxes, Political stability.*

The choice of criteria depends on the freedom of individual action and the elements of harmonization that revolve around integration and the business atmosphere. $\sigma ij$ is a judgment of value appreciated through an evaluation scale or observation. Each criterion has a nature, depending on whether it is an argument "max" or "min".

### 3.2 Choice of resolution method

There are several methods to support multi-criteria decision-making. Although the mathematical tools used to process information can be complex, the basis on which criteria are chosen and performance assessed is often simple, understandable and developed by the group conducting the analysis. The most important advantage of the PROMETHEE method is its ability to simplify complex situations, it allows, by breaking down and structuring the analysis, to precede step by step the search for a solution, in complete transparency.

Through a homogeneous and simultaneous approach to the evaluation of a large number of objects, the method also allows a stable evaluation of the different elements involved in the analysis. We have opted for GAs because they succeed in finding good solutions to very complex problems, and too far from the traditional combinatorial problems to benefit from certain known properties. GAs are robust initialization methods that determine the overall optimum of a functional solution or approach it, and are parallelizable. [3] [4] [7].

The idea in a genetic algorithm is to start from a population of arbitrarily chosen potential, the size of the population is an important factor that has a direct influence on its effectiveness, if the population is composed of a large number of individuals, the probability of obtaining a good quality solution during the process of evolution of the algorithm increases, but the calculation time also increases and makes it less effective, the fitness function is an evaluation of the relative performance of the current solutions of the process.

### 4.3 Initial Population Generation

The effectiveness of genetic algorithms depends on the choice of the initial population that determines the speed of the algorithm, it is necessary to randomly generate individuals by making uniform fingerprints in each of the domains associated with the components of the research space taking into account the constraints of respect by the individuals produced.

**Table 2: List of criteria and actions [Source World Bank 2012 - 2016]**

| Pays | Infrastructure | | Economics | | | Socials | | administrative | | Political |
|---|---|---|---|---|---|---|---|---|---|---|
| | $C_{infr1}$ | $C_{infr2}$ | $C_{eco1}$ | $C_{eco2}$ | $C_{eco3}$ | $C_{soc1}$ | $C_{soc2}$ | $C_{admi1}$ | $C_{admi2}$ | $C_{poli}$ |
| | max | min | max | max | min | max | min | min | max | max |
| Algeria | 1086.3 | 2.44 | 10.73 | -4.36 | 2.7 | 9.88 | 1.8 | 25 | 451 | -1.26 |
| Egypt | 1743.6 | 2.81 | 10.66 | 75.5 | 2.9 | 9.33 | 1.6 | 8 | 426 | -1 |
| Spain | 5636.4 | 3.68 | 0.42 | 227.65 | -0.88 | 21.86 | 0.3 | 34.6 | 186.2 | -0.23 |
| France | 7429.2 | 3.86 | 1.26 | 133.9 | 0.64 | 9.3 | 0.5 | 6.5 | 132 | 0.56 |
| Italy | 5410.2 | 3.66 | 1.33 | 159.45 | -1.04 | 8.5 | 0.4 | 6.8 | 284.4 | 0.47 |
| Libya | 3829.6 | 2.3 | 16.03 | -48 | 1.47 | 4.6 | 1.1 | 30 | 907.6 | -0.25 |
| Morocco | 828.7 | 2.94 | 0.74 | 111.4 | 4.39 | 9.3 | 1.3 | 11.8 | 284.8 | -0.45 |
| Syria | 1677.2 | 2.94 | 0.74 | 49.5 | 3.27 | 8.7 | 2.1 | 14.2 | 336 | -1.26 |
| Tunisia | 1285.3 | 2.99 | 4.94 | 79.03 | 1.74 | 13.5 | 1.2 | 12.5 | 160.8 | -0.19 |
| Turkey | 2712.3 | 3.35 | 6.95 | 69.8 | 4.88 | 9.8 | 1.3 | 6 | 225.1 | -0.99 |

The idea in our approach is to start with a population of potential initial solutions (chromosomes) chosen on the basis of investment location criteria. The GA uses an initial population in the beginning of the evolutionary process that takes place through a sequence of crossing and mutation operators. In this work, the GA begins by randomly generating an initial population of individuals representing solutions that take into account the constraints.

Optimization by Hybridization of a Genetic Algorithm with the PROMOTHEE
Method: Management of Multicriteria Localization
______________________________________________________________

The size of the population is an important factor that has a direct influence on its effectiveness. If the population is composed of a large number of individuals, the probability of obtaining a good quality solution during the evolution process of the algorithm increases, but the computation time also increases and makes it less efficient.

The following algorithm provides the way in which the individual is generated. This generation adopts as a construction strategy solutions that begin with the first level and then move to the second level. It is composed of a set of clusters representing the selection categories. The size of the initial population can be set by a parameter which corresponds to a given percentage of the cardinality of the search space. [4]

It should be noted that, in general, the process of generating the initial population presents a major disadvantage, namely the duplication of individuals. In our case, the risk of having duplicated individuals is reduced by the technique used to produce the population.

Indeed, this method is very effective in ensuring diversity and allowing the convergence of the algorithm. For population selection, we use the selection of the biased wheel that simply chooses parenting solutions based on the value of their fitness to minimize the total cost, the person with the minimum value is more likely to be selected.

### 4.4 Coding of individuals

In evolutionary algorithms, binary coding is most often used. It is adopted according to the nature of the problem to be solved. Although it offers ease of implementation of the various genetic operators, it induces a rather considerable response time.

For our problem, it was adopted the encoding in integers this seems trivial since in the model one seeks, in the end to find the clusters of criteria in the decision of location of the investments. Therefore, the indices of the different nodes of the network will be kept and will be used to represent the individuals. This choice is also due to the fact that it is more convenient to integer coding to designate a large number of individuals and to allow ease of processing, and hence the computational complexity and efficiency are improved. Indeed, the precision of the GA solution encoded in integer numbers is much higher than that of the binary coded GA. The genetic coding of each individual in GA consists of a chain of length 5 * S, where the gene is represented as a cluster which indicates the affection of the categories of the selection criteria.

Myriem Alijo, Otman Abdoun, Mostafa Bachran, Amal Bergam
______________________________________________________________________

```
BEGIN
    Population <-- GenerateInitialPopulation(liste)
        whileGeneration< maximum to EvaluatePopulation(Population)
        BestPopulation<-- Selection(Population)
        CrossPopupation<--Cross(BestPopulation)
        MutatePopuataion<--Mutate(BestPopulation)
        for each pair a and b ∈I we compute (a, b) and (b, a) according to the PROMETHEE
                ifelse (a,b)=0 (a,b)*1/c=(c numbre of critaria)
                we calculate The outgoing flow Φ+
                we calculate The incoming flow Φ-
                we calculate The net flow Φ
        Population<--BestPopulation ∪ CrossPopulation ∪ MutatePopulation

    End while
END
```

**Figure 2. The algorithm of the approach**

### 4. Numerical results
### 4.1 Experimental Testing

We have carried out series of numerical experiments to verify the effectiveness of the proposed resolution approach. Then, this approach was applied for the elaborated model corresponding to the case of location of the future investments.

**5.1.1 Resolution with the PROMETHEE method**

PROMETHEE is a multicriteria decision support method [10][3]. It allows treating multicriteria alternatives for storage. It applies in a process that includes 3 steps:
- Comparing pairs of alternatives from the variables to determine an aggregate preference index;
- calculation of the upgrade flows for the creation of an order on the measurements;
- storage of alternatives from upgrade streams;

If we consider two countries a and b in set I, the over classification of b by *a*, denoted by: "*a* upgrade *b* (*aUb*), is valid if the arguments of a decision maker in favor of the proposition a" is at least As good as b "are sufficiently strong", so if we consider two actions i and k of the set I, the behavior of a decision maker would be such that:

$$\begin{cases} i \text{ is preferred to } k & \forall ik \in I \quad if \quad i \geq k, \delta ij > \delta kj, \; j \in J \\ i \text{ is indifferent to } k & \forall i, k \in I \quad if \quad i \geq k, \delta ij > \delta kj, \; j \in J \end{cases}$$

Optimization by Hybridization of a Genetic Algorithm with the PROMOTHEE
Method: Management of Multicriteria Localization
_______________________________________________________________

The Cartesian product of the set I in I, denoted I x I, is the set of couples (a, b) of the target countries, $\forall a, b \in I$. Let G be the set of pairs (a, b). The elements of each pair are linked by the upgrade relation and G constitutes a graph of the relation. The order through a preference structure and the choices by an upgrade method reflect the existence of a function $\Gamma$ called a "preference function" whose absence of circuits translates a necessary and sufficient condition. [12] [3]

On *m* criteria, we build m preferably unicritorial intensities $\Gamma_j$ *(a, b)* on *(a, b) $\in$ G*. Thus the indicator or intensity of multicriteria preference according to whether the criteria have the same importance $w_j = 1/m$ $\forall j \in J$, such that:

$$\Pi(a,b) = \frac{1}{m} \sum_{j \in J} \Gamma j(a,b)$$

If the criteria do not have the same importance, such that wj (j J) represents the respective weights, then the intensity of preference will be such that:

$$\Pi(a,b) = \sum_{j \in J} wj \Gamma j(a,b)$$

On all the elements of G, we construct the matrix of intensities of multicriteria preference for each pair of actions of I or each element of G. We proceed by comparing two to two actions on a criterion, aggregating the utilities of the pairs of actions related to this criterion. Economically, preference intensities represent an average of the criteria functions that contribute to the preference of one country over another over the assumption of over-classification. The criteria used are true criteria with the same importance, so that each country contributes to the same hypothesis as the others. On all the pairs of the countries of G, the matrix of the intensities of preferences is constructed from the following preference set:

*P (a, b) = {j$\in$ J /$\Gamma$j (a, b) = 1 or 0}.*

This approach to constructing the preference intensity matrix belongs to the family of PROMETHEE methods which takes into account the ex-equation with the risk of having circuits in the upgrade graph. The construction of this matrix assumes the stability of preferences over the countries. A simulation by modifying or adding criteria is appropriate to test the stability of the criteria.

Table 3: Construction of PROMETHEE method overrides relationships

| $\prod$(a,b) | ALGER | EGYPT | SPAIN | FRANCE | ITALY | LIBYA | MAROC | SYRIA | TUNISIA | TURKEY |
|---|---|---|---|---|---|---|---|---|---|---|
| **ALGERIA** | - | 0.5 | 0.6 | 0.7 | 0.6 | 0.5 | 0.6 | 0.5 | 0.5 | 0.5 |
| **EGYPT** | 0.5 |  | 0.4 | 0.5 | 0.7 | 0.5 | 0.5 | 0.6 | 0.5 | 0.4 |
| **SPAIN** | 0.4 | 0.6 | - | 0.4 | 0.7 | 0.6 | 0.6 | 0.6 | 0.6 | 0.6 |
| **FRANCE** | 0.3 | 0.5 | 0.6 |  | 0.6 | 0.5 | 0.6 | 0.5 | 0.4 | 0.5 |
| **ITALY** | 0.4 | 0.3 | 0.3 | 0.4 | - | 0.5 | 0.5 | 0.4 | 0.4 | 0.6 |
| **LIBYA** | 0.5 | 0.5 | 0.4 | 0.5 | 0.5 | - | 0.4 | 0.6 | 0.5 | 0.5 |

| MOROCCO | 0.7 | 0.6 | 0.5 | 0.5 | 0.5 | 0.4 |  | 0.6 | 0.5 | 0.7 |
| SYRIA | 0.5 | 0.4 | 0.4 | 0.5 | 0.6 | 0.4 | 0.6 |  | 0.7 | 0.3 |
| TUNISIA | 0.5 | 0.5 | 0.4 | 0.6 | 0.6 | 0.5 | 0.5 | 0.4 | - | 0.4 |
| TURKEY | 0.5 | 0.6 | 0.4 | 0.5 | 0.4 | 0.5 | 0.5 | 0.7 | 0.6 | - |

These flows aggregate the multicriteria preference intensities of one country relative to the others (outflows) and the multicriteria preference intensities of the others relative to that country (incoming flow). If we take the example of Congo on the matrix Π (a, b), we will have:

$\Phi^+$ (Algeria) = (0. 5 + 0. 6 + 0. 7 + 0. 6 + 0. 5 + 0. 6 + 0. 5 + 0. 5 + 0. 5)/9 = 0.55555556
$\Phi^-$ (Algeria) = (0. 5 + 0. 4 + 0. 3 + 0. 4 + 0. 5 + 0. 4 + 0. 5 + 0. 5 + 0. 5)/9 = 0.44444445
$\Phi$(Algeria) = $\Phi^+$ (Algeria) − $\Phi^-$( Algeria) = 0.1111111

On n = 10 countries, the flow is weighted by n-1 = 9, because of not preferring a country to itself. The following results are obtained for all countries from the matrix Π (a, b):

**Table 4: Numerical Results by using PROMETHEE Method**

| Countries | $\Phi+$ | $\Phi-$ | $\Phi$ | Rang |
|---|---|---|---|---|
| ALGERIA | 0.555555 | 0.444444 | 0.111111 | 5 |
| EGYPT | 0.511111 | 0.488888 | 0.022223 | 6 |
| SPAIN | 0.566666 | 0.433333 | 0.133333 | 2 |
| FRANCE | 0.577777 | 0.411111 | 0.166666 | 1 |
| ITALY | 0.522222 | 0.577777 | -0.055557 | 8 |
| LIBYA | 0.088888 | 0.488888 | -0.300008 | 10 |
| MOROCCO | 0.555555 | 0.433333 | 0.122222 | 3 |
| SYRIA | 0.477777 | 0.522222 | -0.074443 | 9 |
| TUNISIA | 0.488888 | 0.511111 | -0.022223 | 7 |
| TURKEY | 0.533333 | 0.417777 | 0.113555 | 4 |

Before going on to analyze the table, we would like to recall that the PROMETHEE method makes it possible to make a complete classification of the various actions according to the net flows, this method makes it possible to make a total comparison of all the indicators taken into consideration in the process of Territorial decision, the PROMETHEE method is used to make a combination of the incoming flow Φ- and that of the outgoing flows Φ+.

Optimization by Hybridization of a Genetic Algorithm with the PROMOTHEE
Method: Management of Multicriteria Localization
______________________________________________________________

The classification of countries by a total preorder serves to translate the discrepancy of action relative to the choice of the country and the order of the net flows. This ranking makes it possible to measure the competitiveness of countries from the best to the least good. In all the criteria and actions chosen, we note that Libya remains at the bottom of the ranking, followed by Syria, on this chain of preference countries favorable to territorial attractiveness are France, Spain and Morocco.

We have divided the set I into I1 and I2 of the spell that:

*I1 = {France, Spain and Morocco}*
*I2 = {Italy, Syria and Libya}.*

The action of choosing a country for investment location is a multi-decisive multilateral approach. This action poses several challenges for governments to put in place and devise good practices to enhance the attractiveness of different investments. The countries we have classified in sub-group I1 have different policies that help to create favorable conditions and criteria for the economy. We note that the countries in class II have an investment policy supported by the development of infrastructures, from facilitating investment to a good functioning of the administrative system, these indicators were used to elaborate this diagnosis in order to identify the Potential and the strong links of each country. Type *I2* countries have adverse investment outcomes due to local problems, poor governance and weak institutions, and there is a need to boost markets to control Change and to ensure stable inflation.

### 4.2 Hybrid Resolution Approach

This section deals with the hybridization of the genetic algorithm this technique is used to improve the performance of the GA by combining it with another optimization technique knowing which mathematical type of optimization problem we are dealing with, complex or not, is of absolute necessity in IE. It is actually to determine the resolution strategy to follow. A first option is to work on defining analytical resolution procedures, which provide an optimal solution, and a second option is to work on the development of GA, which ensures the optimal nature of the solutions provided.

Myriem Alijo, Otman Abdoun, Mostafa Bachran, Amal Bergam
______________________________________________________________

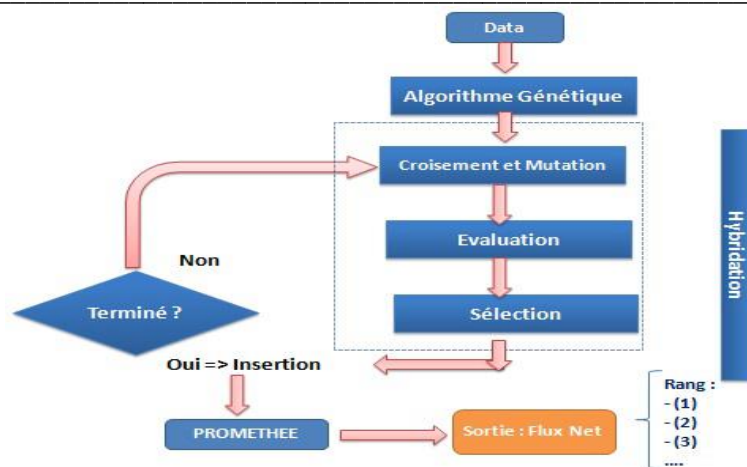

**Figure 3: Organizational chart of the resolution process**

Technically, the complexity of an optimization problem is measured by its difficulty of resolution, which is itself measured by the time required for it to be executed by an information processing system which gives the solution. [1] [2]
The best results are obtained by the following countries:

**Table 5: Classification of performance**

| Country | FluxNet | Rank |
|---|---|---|
| **Country 10** | 0.288889 | 1 |
| **Country 6** | 0.233333 | 2 |
| **Country 4** | 0.1 | 3 |

The best overall performances are obtained respectively by the countries: Countries 10, Countries 6 and Countries 4, which show results that, are clearly superior to the other countries. Our proposed model combines an evolutionary algorithm and multicriteria analysis methods, and digitally certifies the solution with a given precision. The evolutionary algorithm quickly explores the search space to obtain a good solution that can be realized with new criteria and a new ranking in a time n, and then transmits its evaluation to an economic intelligence approach. In this context of the proliferation of initiatives to develop indicators, our model offers potentials that have not yet been explored. The introduction of new hybridization schemes of the indicators makes it possible to classify and apprehend the degree of attractiveness of a territory.

Optimization by Hybridization of a Genetic Algorithm with the PROMOTHEE Method: Management of Multicriteria Localization
______________________________________________________________

**Table 6: Best performance**

| Criteria | Infrastructure | | Economic | | | Social | | Administrative | | Policy |
|---|---|---|---|---|---|---|---|---|---|---|
| | $C_{infra1}$ | $C_{infra2}$ | $C_{eco1}$ | $C_{eco2}$ | $C_{eco3}$ | $C_{soc1}$ | $C_{soc2}$ | $C_{admi1}$ | $C_{admi2}$ | $C_{poli}$ |
| | max | min | max | max | min | max | min | min | max | max |
| Value | 1041.73 | 4.18 | 5.92 | 16.20 | 3.76 | 1.34 | 2.73 | 9.14 | 128.35 | -0.31 |

## 5. Conclusion

The hybridization of the PROMETHEE method and the GA in the economic intelligence process allowed us to rank the countries of the Mediterranean according to their territorial competitiveness obtained from the net and graphical flows of preference. This method is used to rank the countries likely to be the location choice in order of preference from good to bad. The results show that France, Spain and Morocco have strategic advantages to attract foreign investment.

The method also allows us to provide an upgrade chain that highlights two sub-groups that we have named investment-friendly countries (France, Spain, Morocco) and I2 representing countries that are unfavorable to investment (Syria and The Libyans).The I2 sub-group of countries that are unfavorable to investment present obstacles linked to poor governance and to the extraversion that arises in the zone to the crises of dependency thus retaining investment.

We note that our approach is unique and can be carried out over different periods in order to evaluate the efforts made by countries to attract foreign investment and even to evaluate the policies of major projects. The classification of the countries according to our model leads to a modeling of the preferences of the investors in order to better determine their decisional choices of territorial localization.